\documentclass[twocolumn,showpacs,floatfix]{revtex4}
\usepackage{bm}

\ifx\pdfoutput\undefined
\usepackage{graphicx}
\else
\usepackage[pdftex]{graphicx}
\usepackage{epstopdf}
\fi

\usepackage[center]{subfigure}

\begin{document}

\title{Renormalization group
approach to multiscale simulation of
polycrystalline materials using the
phase field crystal model}

\author{Nigel Goldenfeld}
\affiliation{Department of Physics, University of Illinois at
Urbana-Champaign, 1110 West Green Street, Urbana, Illinois, 61801-3080.}
\author{Badrinarayan P. Athreya and Jonathan A. Dantzig}
\affiliation{Department of Mechanical and Industrial Engineering, 1206
West Green Street, Urbana, IL 61801.}

\begin{abstract}

We propose a computationally-efficient approach to multiscale
simulation of polycrystalline materials, based on the phase field
crystal (PFC) model. The order parameter describing the density profile
at the nanoscale is reconstructed from its slowly-varying amplitude and
phase, which satisfy rotationally-covariant equations derivable from the
renormalization group.  We validate the approach using the example of
two-dimensional grain nucleation and growth.

\end{abstract}

%

\maketitle

Why is it so hard to predict the properties of real materials?  Unlike
simple crystalline solids, real materials, produced by a wide range of
processing conditions, contain defects and multiple grains that
strongly impact mechanical, thermal, electrical response, and give rise
to such important phenomena as plasticity, hysteresis, work hardening
and glassy relaxation.  Moreover, it is frequently the case that a
faithful description of materials processing requires simultaneous
treatment of dynamics at scales ranging from the nanoscale up to the
macroscopic.  For example, dendritic growth, the generic mode of
solidification of most metals and alloys, involves the capillary length
at the nanoscale, the emergent pattern dimensions on the scale of
microns, the thermal or particle diffusion length on the scale of
$10^{-4}$m, in addition to the grain and sample size.

Despite these obstacles, progress in rational material design requires a
fundamental understanding of the way in which useful properties
emerge as the mesoscale is approached.  Questions that must be addressed
include: What is the collective behavior of assemblies of nanoscale objects?
How best to achieve target mesoscale properties from nanoscale constituents?
And how can the properties at nano-, meso- and intermediate scales
simultaneously be captured quantitatively and predictively?

A number of computational approaches to handle the range of length scales have
been proposed recently \cite{Phillipsbook, VVED04}, including quasi-continuum
methods\cite{Tadmor, Shenoy, Ortiz, Miller}, the heterogeneous multiscale
method\cite{Weinan1, Weinan2}, multi-scale molecular dynamics\cite{Rudd,
Kaxiras, Robbins, CURT02}, multigrid variants\cite{Fish} and extensions of the
phase field model\cite{warren03}. These techniques strive to provide a unified
description of the many scales being resolved, but in some cases require
non-systematic ways to link the disparate scales to enable treatment of
sufficiently large mesoscale systems. This can introduce spurious modes and
excitations, and difficulties associated with the transition
between scales\cite{VVED04, Weinan2}. Most of this work is limited to
crystalline materials with a few isolated defects\cite{Weinan3}.

In this Letter, we propose a novel theoretical approach to these difficulties,
by combining the phase field crystal (PFC) formalism\cite{ekhg02, eg04} with
renormalization group (RG)\cite{GMOL, NGbook} and related methods (see, e.g.
\cite{BOWM98}), developed for the analysis of hydrodynamic instabilities in
spatially-extended dynamical systems\cite{CGO2, Graham, Nozaki, Sasa, Shiwa,
CN, PN, NPL}.  We present effective equations at the mesoscale, from which the
atomic density can readily be reconstructed, and show that this approach is
capable of generating high fidelity representations of materials processing
dynamics. Moreover,  we demonstrate that the mesoscale equations---analogues of
rotationally-covariant amplitude and phase equations in fluid
convection\cite{Cross}---are computationally tractable and amenable in future
work to adaptive grid techniques.

Our approach is based on a form of the RG which unifies singular perturbation
theory\cite{CGO2}, and is a fully systematic way to extract universal or large
scale structures from spatially-extended dynamical systems.  The basic idea is
to start, not with a molecular dynamics model at the nanoscale, but with a
density functional description (in this context, the phase field crystal
model), whose equilibrium solutions are periodic density modulations.  A system
that is periodic at the nanoscale can be parameterized in terms of a uniform
phase and an amplitude: the amplitude describes the maximum variations in the
density of the system through the unit cell, while the phase describes uniform
spatial translations.  A system with underlying periodicity, but which also
contains defects or other nanostructure, can be represented by a density wave
whose amplitude is at most slowly varying on the nanoscale, and a phase that is
essentially uniform everywhere, except near a defect. This observation suggests
that the phase of the density is the appropriate dynamical variable to use for
describing spatially-modulated nanoscale structure in a mesoscopic system, and
in the vicinity of a defect it must be supplemented by the amplitude.

For any spatially-extended pattern-forming dynamics, RG provides a prescription
for obtaining slowly-varying amplitude and phase equations valid on scales much
larger than the nanoscale\cite{CGO2, Graham, Nozaki, Sasa, Shiwa}.  These
equations possess the key advantage that their solutions are essentially
uniform, with localized rapid variations near defects. The renormalization
procedure used here is more general than real-space renormalization, which has
been attempted in related contexts\cite{Rudd, CURT02}.  In particular, our
technique directly focuses on the instabilities that characterize the dynamics,
which may not have simple real-space interpretations.  Once the amplitude and
phase are determined, the actual structure at the nanoscale (and above) can be
reconstructed.  Because the amplitude and phase equations describe solutions
that are slowly-varying everywhere, except near a defect, adaptive mesh
refinement can be used to solve the amplitude and phase equations.

{\it Phase field crystal model:-\/} The phase field crystal model is a
continuum, nonlinear partial differential equation for the density
$\rho(\vec{x})$ that recently has been shown\cite{ekhg02, eg04} to
capture realistic aspects of materials dynamics, including grain
growth, ductile fracture, epitaxial growth, solidification processes,
and reconstructive phase transitions. In addition, the
model natively supports elasticity theory, both linear and
non-linear, without any {\it ad hoc\/} modeling.  Thus, the PFC can
address the important problem of nanoscale strain effects, and their
coupling from the nanoscale to the continuum.

Let $F\{\rho\}$ denote the coarse-grained free energy functional whose
minima correspond to the equilibrium (lattice) state of a
$d$-dimensional system, and whose corresponding chemical potential
gradient drives the dynamics of $\rho$.  A simple form of $F$ that
gives rise to a triangular lattice equilibrium state is the Brazovskii
form\cite{Brazovskii}:
\begin{equation}
\label{free_energy}
F\{\rho({\bf x})\}=\int d^d\vec{x}\left[ \rho\left(\alpha\Delta T +
\lambda\left(q_o^2+\nabla^2\right)^2\right)\rho/2+u\rho^4/4\right]
\label{eq:Free_energy}
\end{equation}
where $\alpha$, $\lambda$, $q_o$ and $u$ can be related to material
properties\cite{ekhg02,eg04}, and $\Delta T$ denotes the temperature
difference from some (higher) reference temperature.  It is convenient
to rewrite this free energy in dimensionless units, i.e.,
$\vec{x} \equiv \vec{r}q_o$, $\psi \equiv \rho \sqrt{u/\lambda q_o^4}$,
$r\equiv a\Delta T/\lambda q_o^4$, $\tau \equiv \Gamma\lambda q_o^6 t$, where
$\Gamma$ is a phenomenological constant \cite{eg04},
and $F\rightarrow F u/\lambda^2q_o^{8-d}$ so that the equation of
continuity for the density becomes
\begin{equation}
\label{eq:dyn}
\partial \psi/\partial t = \nabla^2\left(\left[r + (1 + \nabla^2)^2\right]
\psi +\psi^3\right)+\zeta.
\end{equation}
\noindent The conserved Gaussian noise
%
will not generally be important for
describing phase transition kinetics, and so will henceforth be
neglected here.

The mean field phase diagram of the PFC equation (\ref{eq:dyn}) can be
calculated analytically\cite{eg04} in a one mode approximation that is
valid in the limit of small $r$, and represented in the plane of
dimensionless temperature, $r$, and dimensionless average density,
$\bar{\psi}$. Three possible equilibrium solutions exist; a `liquid',
$\psi_C = \bar{\psi}$, a two-dimensional `crystal' with triangular
symmetry, $\psi_T = A_T \left(\cos(q_Tx)\cos(q_Ty/\sqrt{3})-
\cos(2q_Ty/\sqrt(3))/2\right)+\bar{\psi}$, and a smectic phase which
will be ignored for present purposes.  The triangular lattice can
exhibit persistent defect structures during the relaxation to equilibrium.

Key to differentiating our approach from others is the fact that the
PFC method was designed to investigate phenomena on diffusive time
scales which are typically many orders of magnitude longer than the
time scales accessible in molecular dynamics simulations. On the other
hand, the PFC method suffers from some of the same drawbacks as
molecular dynamics simulation: by resolving the nanoscale, brute force
computation on a massive scale is required to capture mesoscale
phenomena.  We finesse this difficulty here by working with the
slowly-varying amplitude and phase description.

{\it Mesoscale representation of the PFC:-\/} The dynamics of the
slowly-varying amplitude and phase describes fluctuations about a given
set of lattice vectors, but must be covariant with respect to rotations
of those lattice vectors, in order to properly describe
polycrystalline materials with arbitrarily oriented grains.  A similar
situation arises in describing amplitude and phase variations of
convection rolls, and in the context of the model Swift-Hohenberg
\cite{SH} equations, the form of the governing equations was originally
proposed by Gunaratne et al. \cite{gunaratne}, and derived
systematically from the RG formalism of Chen et al.\cite{CGO2} by
Graham\cite{Graham} (see also ref. \cite{Nozaki}).

The triangular phase solution is represented as
\begin{equation}
\label{1modetriang}
\psi({\vec{x}})=\sum_{j}A_{j}(t)\exp(i{\vec{k_{\it{j}}}\cdot\vec{x}}) +
\bar{\psi},
\end{equation}
where ${\vec{k_{1}}} = k_0 ({-\vec{\mbox{i}}\sqrt{3}/2} -
{\vec{\mbox{j}}/2})$, ${\vec{k_{2}}} = k_0{{\vec{\mbox{j}}}}$ and
${\vec{k_{3}}} = k_0({\vec{\mbox{i}}\sqrt{3}/2} - {\vec{\mbox{j}}/2})$ are the
reciprocal lattice vectors, $k_0$ is the wavenumber of the pattern, $\vec{\mbox{i}}$ and $\vec{\mbox{j}}$ are unit
vectors in the $x$- and $y$-directions, and $A_j$ ($j=1,2,3$) are the complex
amplitude functions. The detailed derivation of the
evolution equations for the RG form of the PFC equation, similar to the
approaches referenced above, will be given elsewhere. We simply present the
result here due to space limitations \begin{equation} \label{rgeqn}
\frac{\partial A_1}{\partial t} = \widetilde{\mathcal{L}}_{1}A -
3A_1\left(|A_1|^2+2|A_2|^2+2|A_3|^2\right) - 6\bar{\psi}A_2^*A_3^*
\end{equation} (together with appropriate permutations for $A_2$ and $A_3$),
where \begin{equation} \label{rcoperator} \widetilde{\mathcal{L}}_{j} =
\left[1 - {\vec{\nabla}}^2 - 2i{\vec{k_{\it{j}}}}\cdot{\vec{\nabla}}\right]
\left[-r-3\bar{\psi}^2- \left\{{\vec{\nabla}}^2 +
2i{\vec{k_{\it{j}}}}\cdot{\vec{\nabla}}\right\}^2\right] \end{equation} is the
manifestly rotationally covariant operator. After solving the RG equations in
Eq. (\ref{rgeqn}), the density is reconstructed using Eq.~(\ref{1modetriang}).

{\it Model Validation:-\/} For ease of numerical implementation,  we have
chosen to solve the amplitude equations about a globally  fixed basis of
lattice vectors. As a result, the grain orientation information must be
borne by the complex amplitude functions $A_j$. We specify this information
through an initial condition $A_j(\theta)$, where $\theta$ is the rotation
angle measured with respect to the basis vectors, and the function
$A_j(\theta)$ is  chosen such that, when the original field $\psi$ is
reconstructed as per Eq. (\ref{1modetriang}), the resulting grain is rotated
by an angle $\theta$.

\begin{figure}[htbp]
\begin{center}
\def\figheight{23mm}
\subfigure[\label{egrg8} $t=56$]
{\includegraphics[height=\figheight,angle=0]{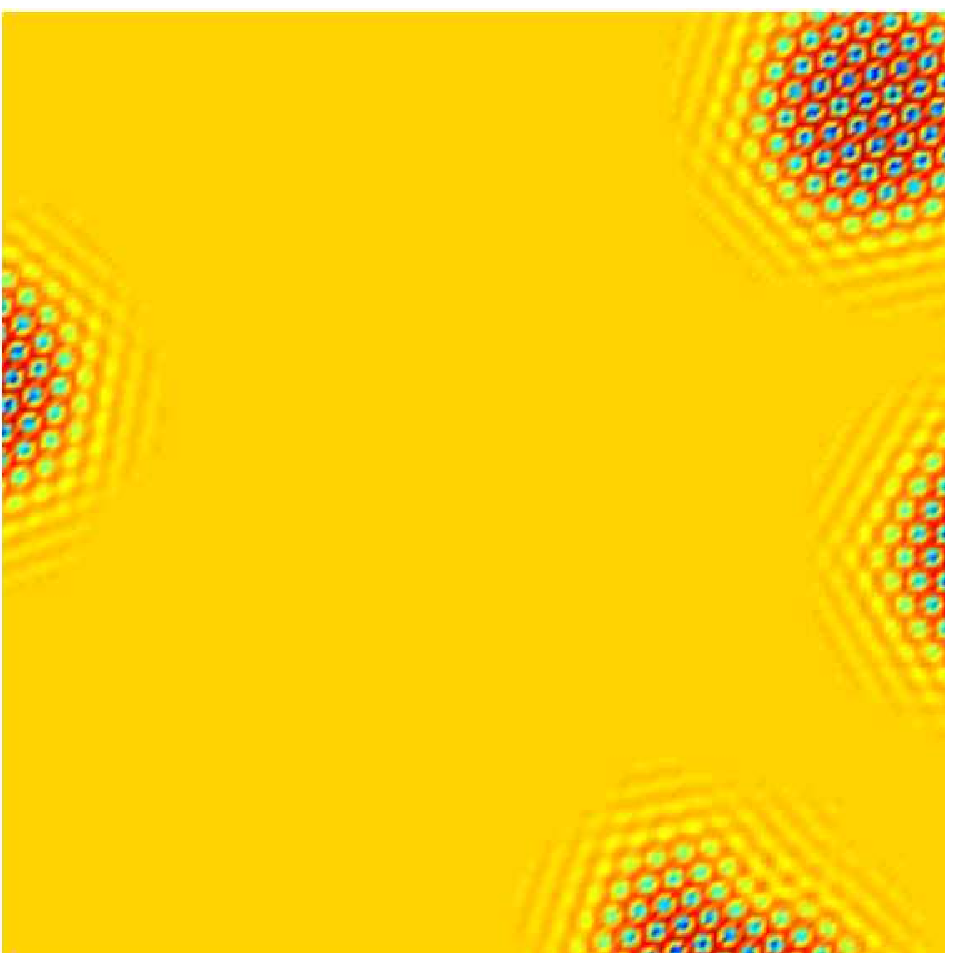}}\hfill
\subfigure[\label{egrg24} $t=184$]
{\includegraphics[height=\figheight,angle=0]{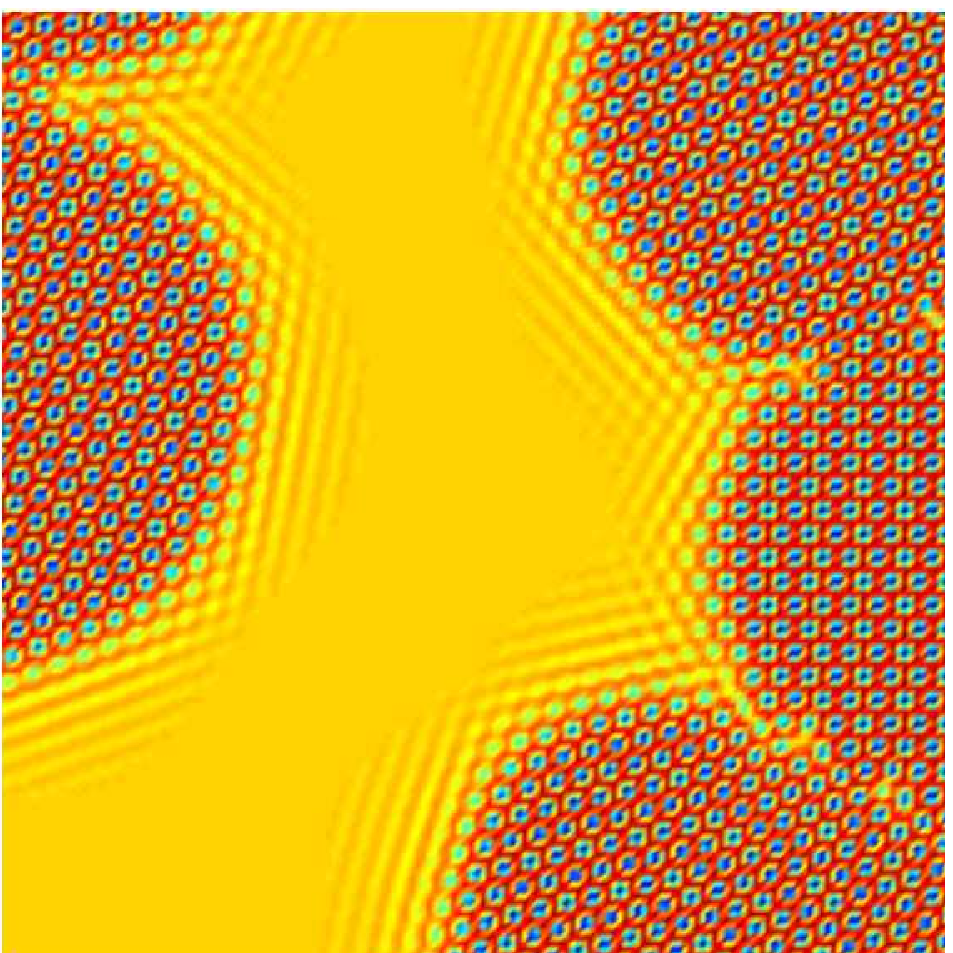}}\hfill
\subfigure[\label{egrg64} $t=720$]
{\includegraphics[height=\figheight,angle=0]{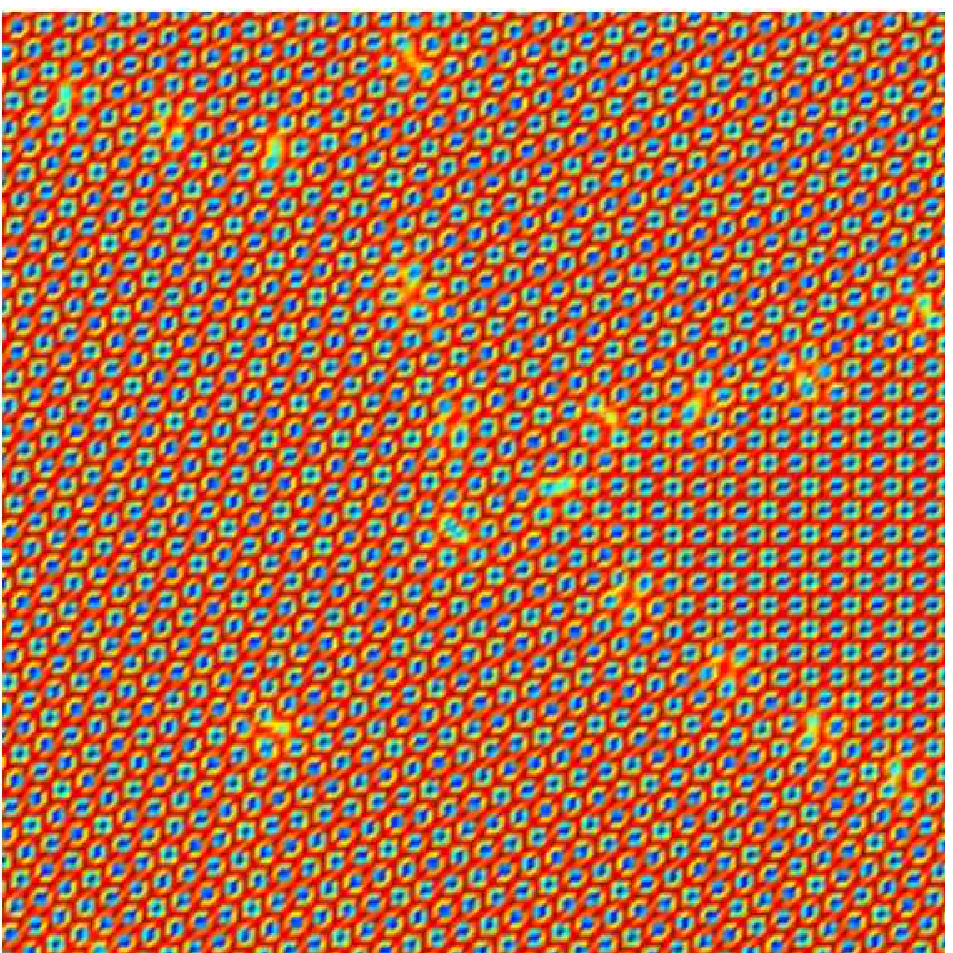}}
\vspace{-5mm}
\caption{\label{fig:comparison}
RG-reconstructed density at indicated times for
heterogeneous nucleation and growth in a 2-D film.}
\end{center}
\end{figure}

Fig.~(\ref{fig:comparison}) shows the time evolution for the nucleation and
growth of a two-dimensional film as predicted by the RG equation, starting
from an initial condition of randomly-oriented seeds, with
$\bar{\psi}=0.285$ and $r=-0.25$. The initial crystallite domains grow,
colliding to form a polycrystalline microstructure.  The solutions obtained
using the PFC equation are essentially indistinguishable from
Fig.~(\ref{fig:comparison}), indicating excellent qualitative agreement. The
key feature of the RG equation is its ability to correctly capture defect
formation and motion.

\begin{figure}[htbp]
\begin{center}
\includegraphics[height=6cm]{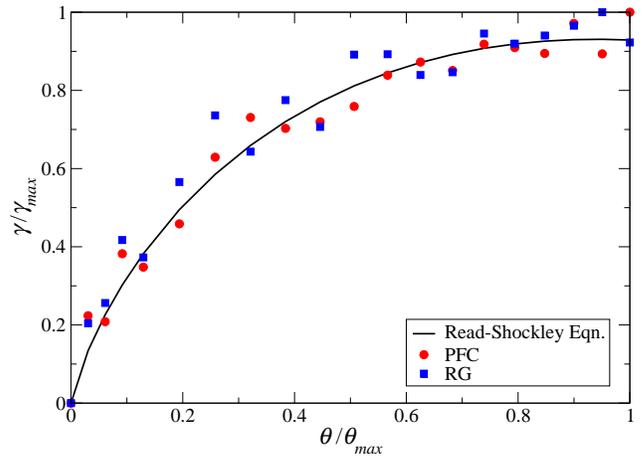}
\caption{\label{ReadShockley} Comparison of grain boundary energy predicted by
the RG and PFC equations, with the Read-Shockley equation.}
\end{center}
\end{figure}

\vspace{-5mm}
As a more rigorous demonstration of accuracy, Fig.~(\ref{ReadShockley})
compares the grain boundary energy, $\gamma$, as a function of the
misorientation angle $\theta$, predicted using the two algorithms. The initial
condition for this test comprised two misaligned crystals separated by a
narrow strip of liquid, on a periodic domain (see \cite{eg04} for details).
Fig.~(\ref{ReadShockley}) also shows the Read-Shockley equation \cite{RS}, a
well known analytical result for small angle grain boundaries, that has been
scaled to fit large misorientation data. The agreement is particularly good
for low angle grain boundaries, and the values predicted by the RG equations
closely follow the trends predicted by the PFC (from \cite{eg04}) and the
Read-Shockley equation. The maximum  difference in the free energy as computed
by the RG and PFC equations is about 1.6\%.

\begin{figure}[htbp]
\begin{center}
\vspace{3mm}
\includegraphics[height=6cm]{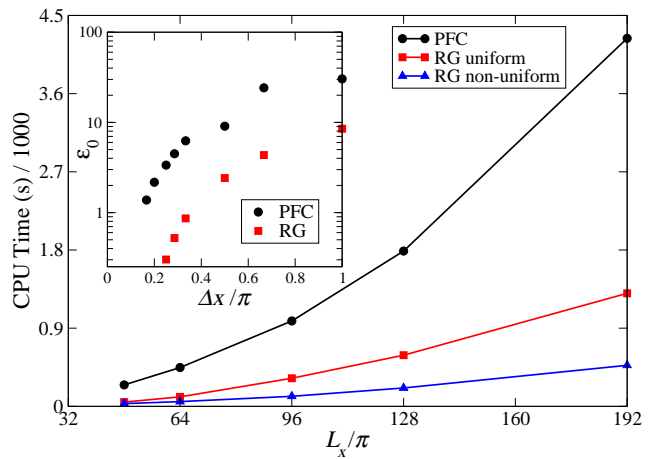}
\caption{\label{speedup} Scaling of CPU time versus domain length $L_x$ for the PFC
and RG equations. Inset shows error in the respective solutions with diminishing mesh
spacing $\Delta x$.}
\end{center}
\end{figure}

{\it Computational efficiency:-\/}  Fig.~\ref{speedup} (inset) shows grid
convergence behavior in the Read-Shockley test of the solutions to the PFC and
RG equations. The crystals are misoriented by the maximum possible angle,
$\pi/6$. We define the error  $\varepsilon_0 = \left|\|y^{\Delta x}\|_2 -
\|y^0\|_2 \right|$,  where $\|y^{\Delta x}\|_2$ is the $L_2$ norm of the
solution for a mesh spacing of $\Delta x$, and $\|y^0\|_2$ is the $L_2$ norm
obtained by Richardson extrapolation to $\Delta x = 0$ consistent with a
second order finite difference method. For a comparable level of accuracy, we
see that $\Delta x_{RG} \approx 2\Delta x_{PFC}$. In a Forward Euler time step
scheme, this leads to a stability condition $\Delta t_{RG} \approx 6\Delta
t_{PFC}$. Clearly, the RG equations offer significant opportunities for
improved computational efficiency.

Fig.~\ref{speedup} compares the CPU time as a function of domain size
$L_x$ for the Read-Shockley test with $\Delta \theta = 3.88^{\circ}$,
$r=-0.25$ and $\bar{\psi}=0.28$, showing that the CPU time required for
the RG equations ranges from about 5 to 6 times less than for the PFC
equations. Consistent with the grid convergence behavior described
above, we chose $\Delta x_{PFC}=\pi/4$, $\Delta t_{PFC}=0.008$, $\Delta
x_{RG}=\pi/2$ and $\Delta t_{RG} = 0.05$. The difference in the free
energy predicted by the RG equations and PFC equations was $<0.1$\%.
Even more significantly, however, the amplitude functions can be solved
on a non-uniform computational grid. For this problem, since the
location of the boundary is known \emph{a priori}, it is easy to
construct an appropriate non-uniform mesh. We chose constant $\Delta
y=\pi/2$, and allowed $\Delta x$ to vary from a minimum of $\pi/2$ to
$2\pi$. This reduced the size of the computational mesh from 257
$\times$ 257 to 97$\times$257. Even with this relatively naive
implementation, we find that the speedup of the RG form compared to the
original PFC form is close to a factor of \emph{ten} (see
Fig.~\ref{speedup}), while the error in the free energy is still
$<0.1$\%.  We anticipate the computational benefits of solving these
equations on a fully adaptive mesh to be much higher.

\begin{figure}[htbp]
\begin{center}
\subfigure[\label{ramp} $A^R_1$]
{\includegraphics[height=3.0cm,angle=0]{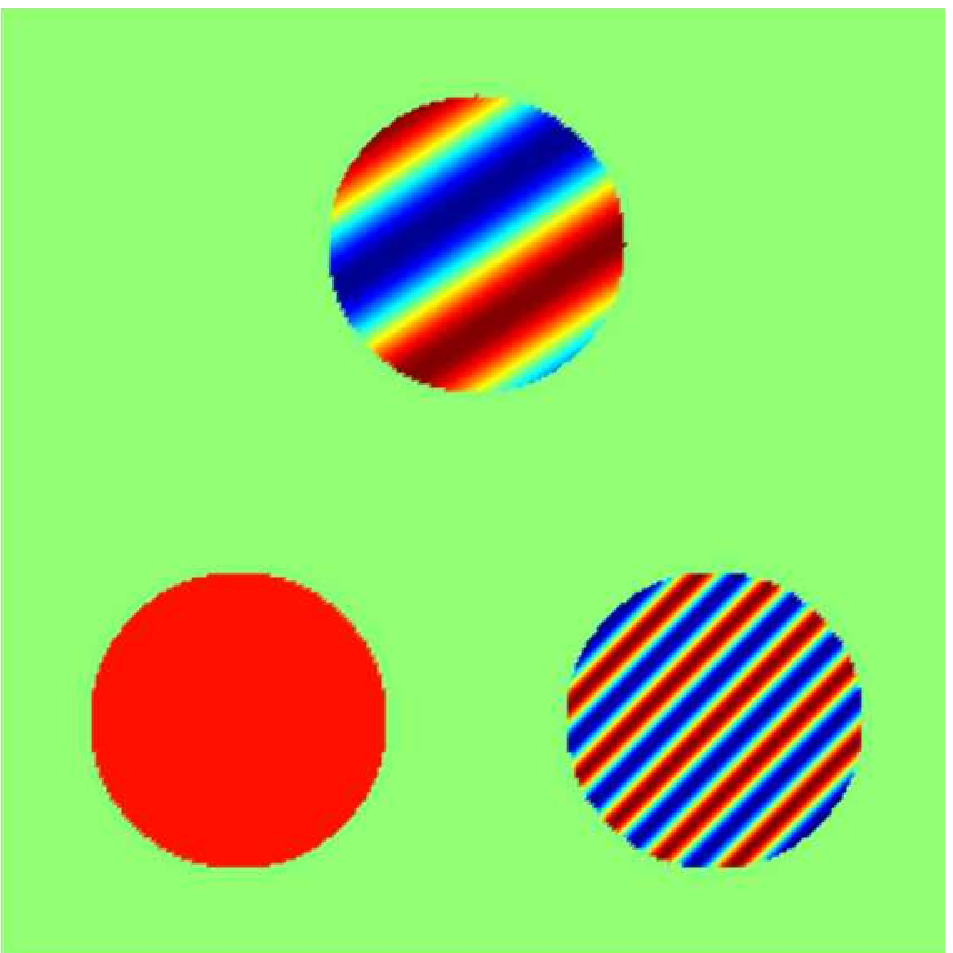}} \hfill
\subfigure[\label{psi} $\psi$]
{\includegraphics[height=3.0cm,angle=0]{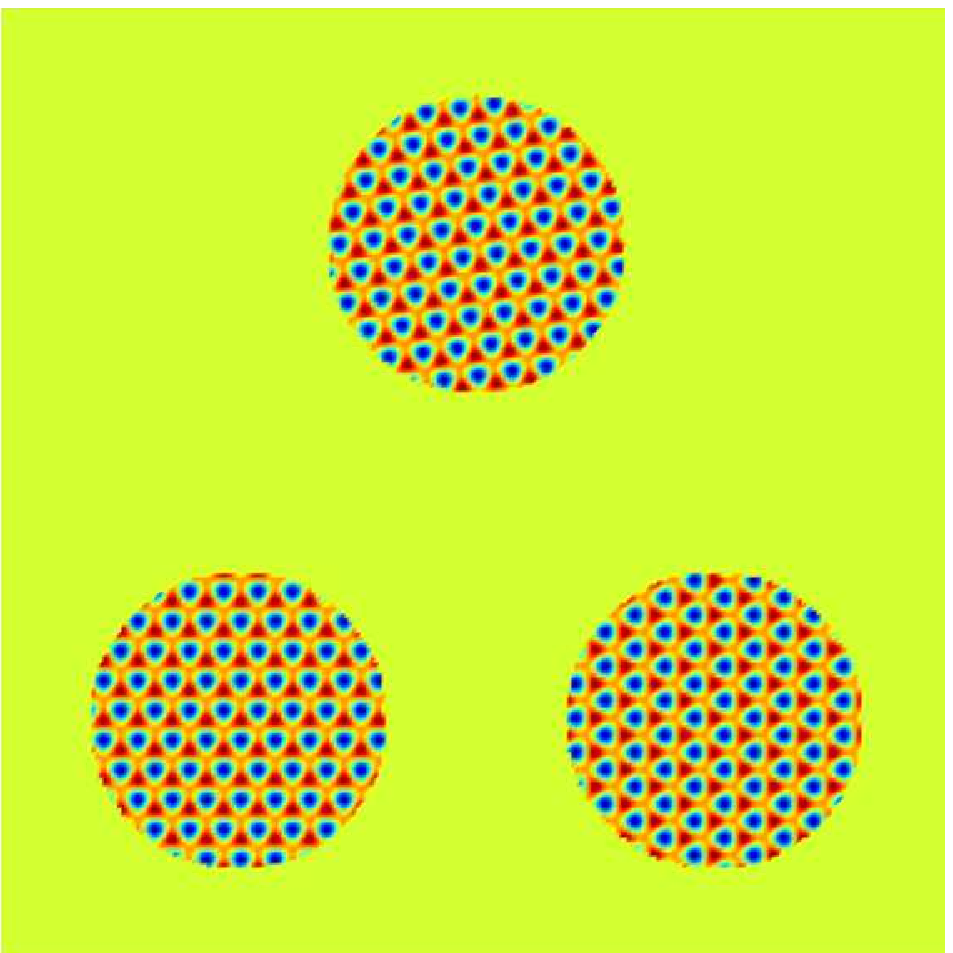}}
\vspace{-5mm}
\caption{\label{fig:beats} (a) Real component of the
complex amplitude; (b) Density field $\psi$
reconstructed using Eq.~(\ref{1modetriang}). Clockwise from the lower left,
$\theta=0$, $\pi/24$ and $\pi/6$.}
\end{center}
\end{figure}

One limitation to the approach described here is that the orientation
information causes a spatial variation in the amplitude when it is represented
using the basis vectors of the triangular lattice. This is illustrated in
Fig.~(\ref{fig:beats}), showing the real part of one of the complex amplitude
functions, and the corresponding reconstructed density variable $\psi$ for
three different orientations.  The ``beats'' evident in Figure \ref{ramp},
which contain grain orientation information, persist as the grains evolve.
This phenomenon could limit the effectiveness of adaptive mesh methods, since
the mesh has to resolve these structures. We note however, that if we write
the complex amplitudes as $A_j = \Psi_j\exp(i\Phi_j)$, where $\Psi_j$ is the
amplitude modulus and $\Phi_j$ is the phase angle, we can formulate equations
of motion for $\Psi_j$ and $\nabla\Phi_j$ from Eq. (\ref{rgeqn}), fields which
are uniform everywhere (no ``beats'') except near defects and interfaces. The
resulting adaptive grid algorithm  can thus be made to scale much more
optimally \cite{PGD}, with interface/grain boundary length rather than the
area of significantly misoriented grains. We will present this work in a
future article.

In summary, we have shown that multiscale modeling of complex polycrystalline
materials microstructure is possible using a combination of continuum modeling
at the nanoscale using the PFC model, RG and related techniques from
spatially-extended dynamical systems theory. Our equations are well-suited
for efficient adaptive mesh refinement, thus enabling realistic modeling of
large-scale materials processing and behavior.

This work was supported in part by the National Science Foundation through grant
NSF-DMR-01-21695 and by the National Aeronautics and Space Administration through
grant NAG8-1657. We thank Prof. Ken Elder for several useful discussions.

\bibliographystyle{apsrev}
\bibliography{./multiscale}

\end{document}